\documentclass[letterpaper, 10 pt, conference]{ieeeconf}
\usepackage[utf8]{inputenc}

\IEEEoverridecommandlockouts    
\usepackage{cite}
\usepackage{amsmath,amssymb,amsfonts,amstext}
\usepackage{graphicx}
\usepackage{textcomp}

\usepackage[utf8]{inputenc}
\usepackage{babel}
\usepackage{balance}
\usepackage{comment}
\usepackage[position=bottom]{subfig}

\usepackage{mathtools}
\usepackage{float}

\usepackage{amsthm}

\newtheorem{problem}{Problem}

\usepackage{algorithm}
\usepackage{algorithmicx} 
\usepackage{algpseudocode} %

\usepackage{todonotes}

\makeatletter
\let\NAT@parse\undefined
\makeatother
\usepackage{hyperref}
\hypersetup{
  colorlinks=true,
    linkcolor= blue,
    allcolors=blue,
    citecolor = blue,
    filecolor=black,      
    urlcolor=blue,
    }

\title{\bf Route Recommendations for Traffic Management\\Under Learned Partial Driver Compliance}

\author{Heeseung Bang$^1$$^*$, \textit{Member, IEEE}, Jung-Hoon Cho$^2$$^*$, \textit{Graduate Student Member, IEEE}, \\
    Cathy Wu$^2$, \textit{Member, IEEE}, and Andreas A. Malikopoulos$^1$, \textit{Senior Member, IEEE}
    \thanks{This research was supported in part by NSF under Grants CNS-2401007, CMMI-2348381, IIS-2415478, CNS-2149548, and in part by MathWorks.}
    \thanks{$^*$ Equal contribution}
	\thanks{$^1$ Heeseung Bang and Andreas Malikopoulos are with the School of Civil and Environmental Engineering, Cornell University, Ithaca, NY 14850, USA. {\tt\small email: \{h.bang,amaliko\}@cornell.edu}}
	\thanks{$^2$ Jung-Hoon Cho and Cathy Wu are with the Department of Civil and Environmental Engineering, Massachusetts Institute of Technology, Cambridge, MA 02142, USA. {\tt\small email: \{jhooncho,cathywu\}@mit.edu}}
}

\date{March 2025}

\begin{document}

\maketitle
\begin{abstract}
In this paper, we aim to mitigate congestion in traffic management systems by guiding travelers along system-optimal (SO) routes. However, we recognize that most theoretical approaches assume perfect driver compliance, which often does not reflect reality, as drivers tend to deviate from recommendations to fulfill their personal objectives. Therefore, we propose a route recommendation framework that explicitly learns partial driver compliance and optimizes traffic flow under realistic adherence. We first compute an SO edge flow through flow optimization techniques. Next, we train a compliance model based on historical driver decisions to capture individual responses to our recommendations. Finally, we formulate a stochastic optimization problem that minimizes the gap between the target SO flow and the realized flow under conditions of imperfect adherence. Our simulations conducted on a grid network reveal that our approach significantly reduces travel time compared to baseline strategies, demonstrating the practical advantage of incorporating learned compliance into traffic management.

\end{abstract}

\section{Introduction} \label{sec:introduction}

Traffic congestion remains a critical challenge in large-scale transportation networks \cite{Chremos2020MechanismDesign}, imposing substantial economic \cite{arnott1994economics}, environmental \cite{chin1996containing}, and social costs \cite{walters1961theory,mao2012social}.
Despite significant investments in transportation infrastructure and advances in traffic management systems, the efficient utilization of existing road networks remains constrained by the complex interaction between system-level optimization and individual driver behavior. 
Common approaches, such as dynamic traffic assignment, are fundamentally rooted in system-level optimization and frequently assume perfect driver compliance or simplified behavior models \cite{bang2024emergingequity}.
For instance, modern navigation tools typically offer routes based on individual travel time minimization. However, this self-focused behavior can amplify congestion and deviate significantly from the \textit{system optimal} (SO) distribution, which minimizes total network travel time by coordinating route choices.

While many theoretical models compute SO flows, their effectiveness hinges on full driver compliance—an assumption rarely met in practice. Real drivers often deviate to fulfill personal objectives such as shorter travel times, lower tolls, or route familiarity. This gap between theoretical SO assignments and real-world decision-making underscores the need for strategies that accommodate partial driver adherence.
Recent research has begun exploring the middle ground between these approaches.

For instance, congestion-aware routing in autonomous mobility-on-demand systems uses dynamic programming or model predictive control to balance route assignments and rebalancing tasks \cite{salazar2019congestion,Bang2022combined,wollenstein2020congestion,bang2021AEMoD}. More recently, reinforcement learning techniques have been introduced to address route guidance under varying degrees of compliance or uncertain traffic conditions \cite{mushtaq2023multi,jiang2024regional,yun2024navigating}. 
While these studies take important steps toward adaptive routing, many either investigate incentive programs to promote desired behavior \cite{niazi2024incentive} or how route recommendations might influence driver behavior to improve system-level outcomes \cite{dia2007modelling}.
However, these efforts typically make simplistic assumptions about driver compliance, treating it as either deterministic or governed by basic probability distributions that fail to capture the nuanced decision-making processes of actual drivers \cite{letchner2006trip,bao2022modeling,zhou2024expert}, while it depends on numerous factors, including perceived recommendation credibility, detour magnitude, traffic conditions, and individual preferences.

In this paper, we introduce a route recommendation framework that bridges the gap between SO traffic assignment and practical implementation by explicitly modeling and learning driver compliance patterns. Our approach first computes the macroscopic flow distribution that optimizes overall system performance. Rather than assuming perfect compliance, we use empirical data to learn how drivers respond to different types of route recommendations. This learned compliance model informs a stochastic optimization problem to minimize the discrepancy between SO flow and expected flow resulting from driver responses to our recommendations.

The contributions of our paper are multifaceted and significant to the domain of traffic management. First, we present a mathematical formulation for compliance-aware route recommendations that effectively bridges the gap between system optimization and behavioral realities. Second, we introduce a data-driven approach for modeling driver compliance patterns, thereby enhancing our understanding of individual driver behavior. Additionally, we develop a stochastic optimization framework that accommodates the inherent uncertainty in driver responses, thereby improving the robustness of our recommendations. Furthermore, we provide numerical evidence demonstrating the efficacy of our approach across a variety of network configurations and demand scenarios. By directly addressing the critical implementation gap in traffic management, our work offers practical pathways toward more efficient utilization of existing traffic infrastructure. Importantly, we acknowledge the fundamental role of human behavior within complex socio-technical systems \cite{chremos2020MobilityMarket}, ensuring that our recommendations are not only theoretically sound but also practically applicable in real-world contexts.

\section{System Architecture} \label{sec:structure}

\begin{figure}
    \centering
    \includegraphics[width=0.85\linewidth]{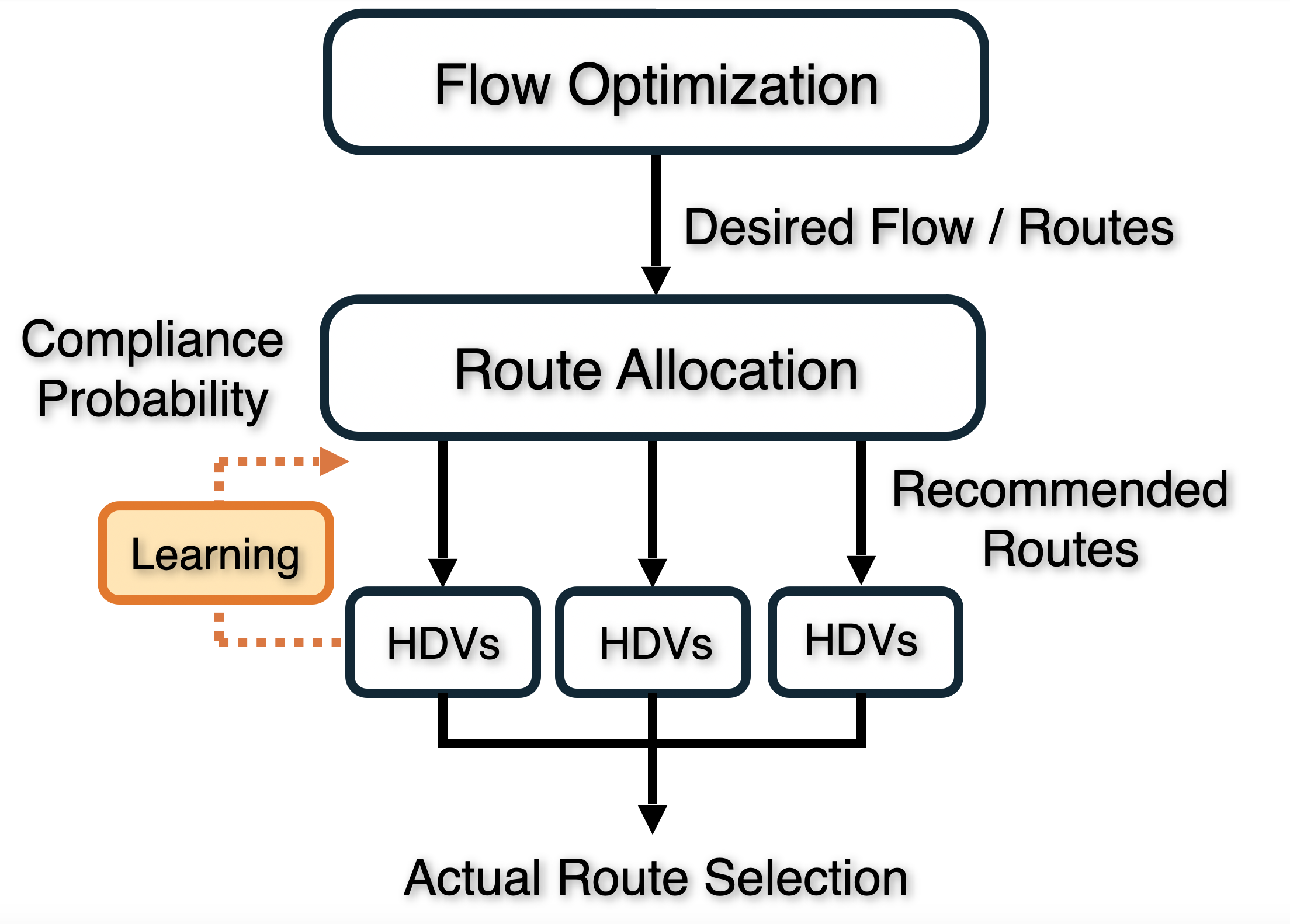}
    \caption{System-optimal (SO) route recommendation framework under partial driver compliance.}
    \vspace{-1em}
    \label{fig:structure}
\end{figure}

This section presents our system architecture for compliance-aware route recommendations in a traffic network.
Figure \ref{fig:structure} illustrates the structure of our approach, which incorporates driver compliance behavior and SO flow into the recommendation process.
The framework comprises several interconnected components that generate route recommendations that balance system-level efficiency with the realities of driver behavior.

The framework begins with a flow optimization component that computes the SO flow distribution across the road network. This optimization takes as input the network topology, travel demand rates, and road characteristics to determine the theoretical flow pattern that minimizes total system travel time if all drivers follow assigned routes perfectly. The resulting optimal flow, however, may not be directly implementable due to unexpected route choices by individual drivers.
To address this challenge, we use machine learning to predict how drivers respond to route suggestions based on historical driving data. The learning model would handle contextual features such as trip characteristics and traffic conditions, and eventually, allow us to anticipate drivers' behavior.

The route recommendation component represents the core of our framework, where SO flows meet behavioral realities. This component takes as input both the desired SO flow pattern and the learned compliance models to generate personalized route recommendations for individual drivers. By formulating a stochastic optimization problem that minimizes the discrepancy between the SO state and the expected state resulting from driver decisions, this component produces recommendations that are both aligned with system-level objectives and informed by observed driver behavior patterns.


To this end, we provide a formal mathematical formulation of each component of this framework, beginning with the problem of computing SO flows and then proceeding to the compliance modeling and route recommendation components that form the core of our contribution.

\section{Problem Formulation} \label{sec:problem}

Consider a traffic network given by a directed graph $\mathcal{G} = (\mathcal{V},\mathcal{E})$ where $\mathcal{V}\subset\mathbb{N}$ is a set of nodes representing intersections (or junction points) and $\mathcal{E}\subset\mathbb{N}$ is a set of edges representing roads.
Let $\mathcal{M}\subset\mathbb{N}$ denote a set of travel demands, where each travel demand $m\in\mathcal{M}$ is characterized by an origin node $o_m\in\mathcal{V}$, a destination node $d_m\in\mathcal{V}$, and a demand rate $\alpha_m\in\mathbb{R}_{\geq0}$, representing the number of travelers per unit of time.
For each travel demand $m\in\mathcal{M}$, we consider a set of possible paths $\mathcal{P}_m$, where each path $p\in\mathcal{P}_m$ is a sequence of connected edges from origin $o_m$ to destination $d_m$.
We denote the union of all path sets as $\mathcal{P}=\bigcup_{m\in\mathcal{M}} \mathcal{P}_m$.

\subsection{Flow optimization}

The vehicle flow on each path $p\in\mathcal{P}_m$ is denoted by $x_p\in\mathbb{R}_{\geq 0}$, and the vehicle flow $x_e$ on each edge $e\in\mathcal{E}$ can be derived from the path flows, i.e.,
\begin{equation}
    x_e=\sum_{m\in\mathcal{M}}\sum_{p\in\mathcal{P}_m} x_p \cdot \mathbb{I}(e\in p),
\end{equation}
where $\mathbb{I}(e\in p)$ is an indicator function that yields $1$ if edge $e$ is part of path $p$ and $0$ otherwise.

To capture the relationship between traffic flow and travel time, we employ the widely-used Bureau of Public Roads (BPR) latency function \cite{us1964traffic}.
The travel time on the road $e\in\mathcal{E}$ with flow $x_e$ is given by
\begin{equation}
    t_e(x_{e}) = t_{e}^0\left(1+0.15\left(\frac{x_{e}}{\gamma_{e}}\right)^4\right), \label{eqn:BPR}
\end{equation}
where $t_{e}^0 \in\mathbb{R}_{>0}$ is the free-flow travel time and $\gamma_e \in\mathbb{R}_{>0}$ is capacity of the road $e\in\mathcal{E}$.
This captures the non-linear relationship between traffic flow and congestion, where travel time increases more rapidly as flow approaches and exceeds capacity.
Next, we formulate vehicle-flow optimization problem.

\begin{problem}[Flow-based routing] \label{prb:flow_routing}
We aim to find the SO vehicle flow distribution by solving the following:
\begin{equation}
\begin{aligned}
    \min_{\mathbf{x}} ~&J(\mathbf{x}) = \sum_{e\in\mathcal{E}} \bigg\{ t_{e}(x_{e})\cdot x_{e}\bigg\} \label{eqn:flow_routing}\\
    \text{s.t. } & \sum_{p\in\mathcal{P}_m} x_p = \alpha_m,~\forall m\in\mathcal{M},\\
    & x_p \geq 0,~\forall p\in\mathcal{P},
\end{aligned}    
\end{equation}
where $\mathbf{x} = (x_p)_{p\in\mathcal{P}}$ is the vector of all path flows.
\end{problem}
The objective function $J(\mathbf{x})$ represents the total system travel time, and minimizing this function yields the SO flow distribution.
The first constraint ensures that the total flow across all paths for each travel demand $m\in\mathcal{M}$ satisfies the demand rate $\alpha_m$, while the second constraint enforces non-negative flows.
This formulation provides the theoretical optimal flow pattern that would minimize overall system congestion. However, achieving this optimal flow in practice requires addressing driver compliance behaviors, which will be the focus of the following sections.

\subsection{Compliance Probability Estimation}
To model driver compliance with route recommendations, we utilize a data-driven approach based on historical observations.
Let $\mathcal{D}=\{(z_n, p_n^\mathrm{r}, y_n)\}_{n=1}^{\mathbf{N}}$ denote a dataset of $\mathbf{N}$ past observations, where $z_n \in \mathbb{R}^d$ is a feature vector describing the $n$-th travel scenario and the corresponding traveler and $p_n^\mathrm{r}$ is recommended path selected from the candidate path set $\mathcal{P}_n$, where $\lvert\mathcal{P}_n\rvert > 2$.
The feature vector $z_n$ typically includes information such as the traveler's origin and destination, demographic characteristics, and other contextual factors that travelers have consented to share with the system operator. The label $y_n \in \{0,1\}$ indicates whether the traveler complied ($1$) or deviated ($0$) from the recommended path during that observation.

Using this dataset, we employ a supervised-learning approach to estimate a function $\phi:\mathbb{R}^{d+1} \to [0,1]$ that predicts the probability of compliance given the features $z$ and recommend path $p^\mathrm{r}$. Formally, the goal is to solve
\begin{equation}
    \hat{\phi} =\underset{\phi\,\in\,\Phi}{\mathrm{arg\,min}}\frac{1}{\mathbf{N}}\,\sum_{n=1}^\mathbf{N} \ell\bigl(y_n,\,\phi(z_n, p_n^\mathrm{r})\bigr),
\end{equation}
where $\ell(\cdot,\cdot)$ is a loss function (e.g., cross-entropy), and $\Phi$ is a class of permissible predictors. The learned predictor $\hat{\phi}$ then outputs a compliance probability in $[0,1]$ for any new scenario $z$ and recommended path $p^\mathrm{r}$.

A Random Forest (RF) \cite{breiman2001random} is a popular choice for the function class $\Phi$ due to its robustness and ability to capture complex relationships in the data.
RF is an ensemble of $T$ decision trees, each grown by recursively partitioning the feature space into leaf nodes. The $t$-th decision tree produces an estimated probability $h_{t}(z, p^\mathrm{r})$ of compliance, often computed by the fraction of training samples labeled $y_n=1$ in the leaf node where the input $z$ and $p^\mathrm{r}$ land. 
After training, the model provides, for each new feature vector $z$ and recommended path $p^\mathrm{r}$, a probability $\hat{\phi}(z,p^\mathrm{r})\in [0,1]$ that quantifies the estimated likelihood of compliance:
\begin{equation}
    \hat{\phi}(z,p^\mathrm{r}) = \frac{1}{T}\,\sum_{t=1}^{T} h_{t}(z, p^\mathrm{r}).
\end{equation}

\subsection{Route Recommendation}
For each traveler $n\in\mathcal{N}$, we define a compliance probability function $\phi_n(p~|~p_n^\mathrm{r})$ that represents the probability of traveler $n$ choosing path $p$ when recommended path $p_n^\mathrm{r}$. In practice, however, the probability of a traveler choosing an alternative option (i.e., $p \neq p_n^\mathrm{r}$) cannot be directly estimated because such events are not explicitly observed in the training data. Consequently, we assume that if a traveler does not follow the recommended path, the probability of selecting any one of the unrecommended options is uniform across all alternatives. 
This can be modeled as
\begin{equation}
    \phi_n\bigl(p \mid p_n^\mathrm{r},z_n\bigr)= 
   \begin{cases}
       \hat{\phi}_n\bigl(z_n,p_n^\mathrm{r}\bigr), 
         & \text{if } p = p_n^\mathrm{r}, \\
       \frac{1-\hat{\phi}(z_n, p_n^\mathrm{r})}{|\mathcal{P}_n|-1}, 
         & \text{if } p \neq p_n^\mathrm{r}.
   \end{cases}
\end{equation}
For simplicity of notation, we write $\phi_n\bigl(p \mid p_n^\mathrm{r}\bigr)$ instead of $\phi_n\bigl(p \mid p_n^\mathrm{r}, z_n\bigr)$, with the understanding that this probability may still depend on the underlying features $z_n$.

To translate the SO flow $x_e^*$ on each road $e \in \mathcal{E}$ into a particular measure, we apply Little's Law \cite{little1961proof} to approximate the number of vehicles $L_e^*$ on the road $e$:
\begin{equation}
    L_e^* = x_e^* \cdot t_{e}(x_e^*), \quad \forall e \in \mathcal{E},
 \end{equation}
where $t_e(x_e^*)$ is the average travel time on edge $e$ as given by \eqref{eqn:BPR}. This captures the steady-state relationship between vehicle flow and travel time on each edge.

Given the SO occupancy $L_e^*$ and our compliance probability model, we formulate a route recommendation problem that aims to minimize the discrepancy between the desired SO occupancy and the expected occupancy resulting from drivers' decision:

\begin{problem}[Route Allocation] \label{prb:assignment} The route allocation problem is formulated as follows:
    \begin{equation}
        \min_{\{p_n^\mathrm{r}\}} \sum_e \Bigl|L_e^* - \sum_n\sum_p \phi_n(p|p_n^\mathrm{r}) \cdot \mathbb{I}(e\in p)\Bigr|.
    \end{equation}
\end{problem}

Problem \ref{prb:assignment} seeks the set of route recommendations $\{p_n^\mathrm{r}\}$ that minimizes the difference between the SO occupancy and the expected actual occupancy across all network edges while accounting for probabilistic compliance behavior.

\section{Solution Approach and Simulations} \label{sec:simulation}

This section provides the solution approaches for each component of our framework and evaluates their performance through numerical simulations.
We first present the computational methods used to solve the vehicle-flow optimization problem and describe our implementation and analysis of the machine learning pipeline for compliance probability estimation.
We then explain the integer programming solution approach for Problem~\ref{prb:assignment} and evaluate our framework with comparative scenarios.

\subsection{Network and Data}
The traffic network is given by a directed graph $\mathcal{G} = (\mathcal{V},\mathcal{E})$, as defined above.
In the simulation examples, the graph $\mathcal{G}$ is chosen to be a grid with $R \times C$ nodes, although any directed network can be employed in principle. Each road $e\in\mathcal{E}$ is assigned a length $\ell_{e}\in\mathbb{R}_{\ge0}$, capacity $u_{e}\in\mathbb{R}_{\ge0}$, risk $r_{e}\in[0,1]$, and toll $c_{e}\in\mathbb{R}_{\ge0}$. The free-flow travel time on $e$ is $t_{0,e}\in\mathbb{R}_{\ge0}$, and a maximum travel time $t_{e}^{\mathrm{max}}\in\mathbb{R}_{\ge0}$ is included for normalization. The function $t_{e}(x_{e})$ follows the common BPR form so that increased flow $x_{e}$ raises the travel time on edge $e$.

To model individual route choices, each demand $m\in\mathcal{M}$ can be decomposed into discrete travelers or agents who share the same origin $o_m$ and destination $d_m$. Each traveler $i$ of type $m$ has a cost function that includes components for road risk, travel time, and tolls. In addition, the cost model incorporates an adherence term linked to a recommended path and may include incentives to reduce financial burdens. Specifically, the total cost of a path $p$ for traveler $i$ is
\begin{align*}
    J_{p,i} &=\sum_{e\,\in\,p} \Bigl(\theta_{1,i}\,r_{e} +\theta_{2,i}\,\tfrac{t_{e}(x_{e})}{t_{e}^{\mathrm{max}}}+\theta_{3,i}\,c_{e}\Bigr)+\theta_{4,i}\,\mathbb{I}\{p \neq p^{\mathrm{r}}\}.
\end{align*}
The notation $\theta_{1,i}$, $\theta_{2,i}$, $\theta_{3,i}$, and $\theta_{4,i}$ reflects the traveler's preferences regarding road risk, time, monetary costs, and adherence to the recommended path, respectively. 
$\mathbb{I}\{p \neq p^{\mathrm{r}}\}$ is an indicator function that adds an adherence-related penalty if the traveler deviates from the recommended path.

In the numerical studies, these preference weights arise from latent parameters sampled according to each traveler's personal or demographic characteristics. After normalization of the latent parameters, the resulting weights capture, for instance, high or low risk tolerance and strong or weak sensitivity to travel time. The choice of the final path follows a Boltzmann (softmax) rule, where the probability of selecting path $p$ from a finite set of candidates is proportional to $\exp(-\lambda J_{p,i})$ for a rationality parameter $\lambda>0$. This probabilistic selection allows for realistic variability in route choices, rather than assigning all travelers to their absolute minimum-cost path.

The simulation proceeds by generating the underlying base flow on each road, assigning a recommended path $p^{\mathrm{r}}$ for each traveler, and then sampling the traveler's final path according to the softmax probabilities. Once all travelers have selected paths, the corresponding flows are tallied to update $x_{e}$. The primary outputs are the traveler-level decisions (e.g., compliance rates, chosen paths, cost) and the resulting edge-level flows. Both full and partial compliance are tracked; the latter is measured by comparing the overlap between the recommended path and the traveler’s chosen path.

\subsection{Flow optimization}

\begin{figure}
    \centering
    \includegraphics[width=0.72\linewidth]{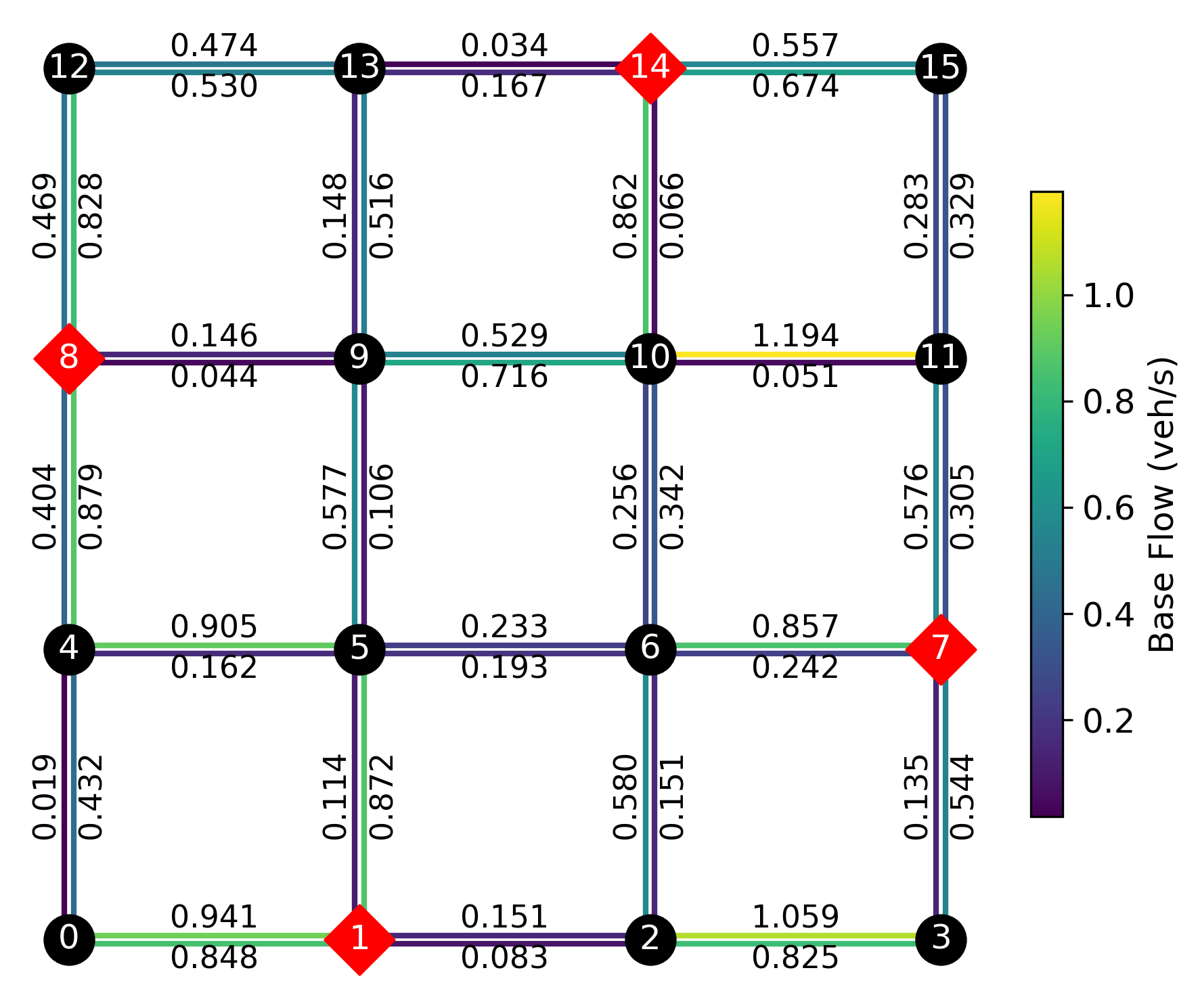}
    \caption{Example $4\times4$ grid network and its base flow distribution. Nodes marked in red serve as designated origins and destinations. Each edge label indicates the base flow rate.}
    \vspace{-1.2em}
    \label{fig:network-baseflow}
\end{figure}

For the flow optimization, we utilize a $4\times4$ grid network with a randomly generated base flow as illustrated in Fig. \ref{fig:network-baseflow}.
This network represents a simplified urban road network.
We selected four nodes ($1$, $7$, $8$, $14$) to be origins and destinations with $12$ distinct Origin-Destination (OD) pairs. Each OD pair has a travel demand of $0.33$ vehicles per second ($1200$ vehicles per hour), which is moderate traffic conditions in urban environments.
To account for the existing base flows in the network, we modify the objective function to become $J(\mathbf{x}) = \sum_{e\in\mathcal{E}} \left\{t_e(x_e+f_e)\cdot x_e\right\}$ where $f_e$ represents the base flow on edge $e\in\mathcal{E}$.
Then, we solve Problem \ref{prb:flow_routing} using sequential least squares programming in SciPy \cite{2020SciPy-NMeth}.


\subsection{Learning Compliance Probability}
We collected data over 200 days, each containing records of driver recommendations and subsequent route choices. We partitioned this dataset into three subsets—training (60\%), validation (20\%), and evaluation (20\%)—ensuring adequate representation of various origins, destinations, and route attributes in each split. An RF regressor was trained on the training set and tuned using the validation set. Finally, we evaluated its performance on the remaining test set.

\begin{figure}[!t]
    \centering
    \includegraphics[width=0.75\linewidth]{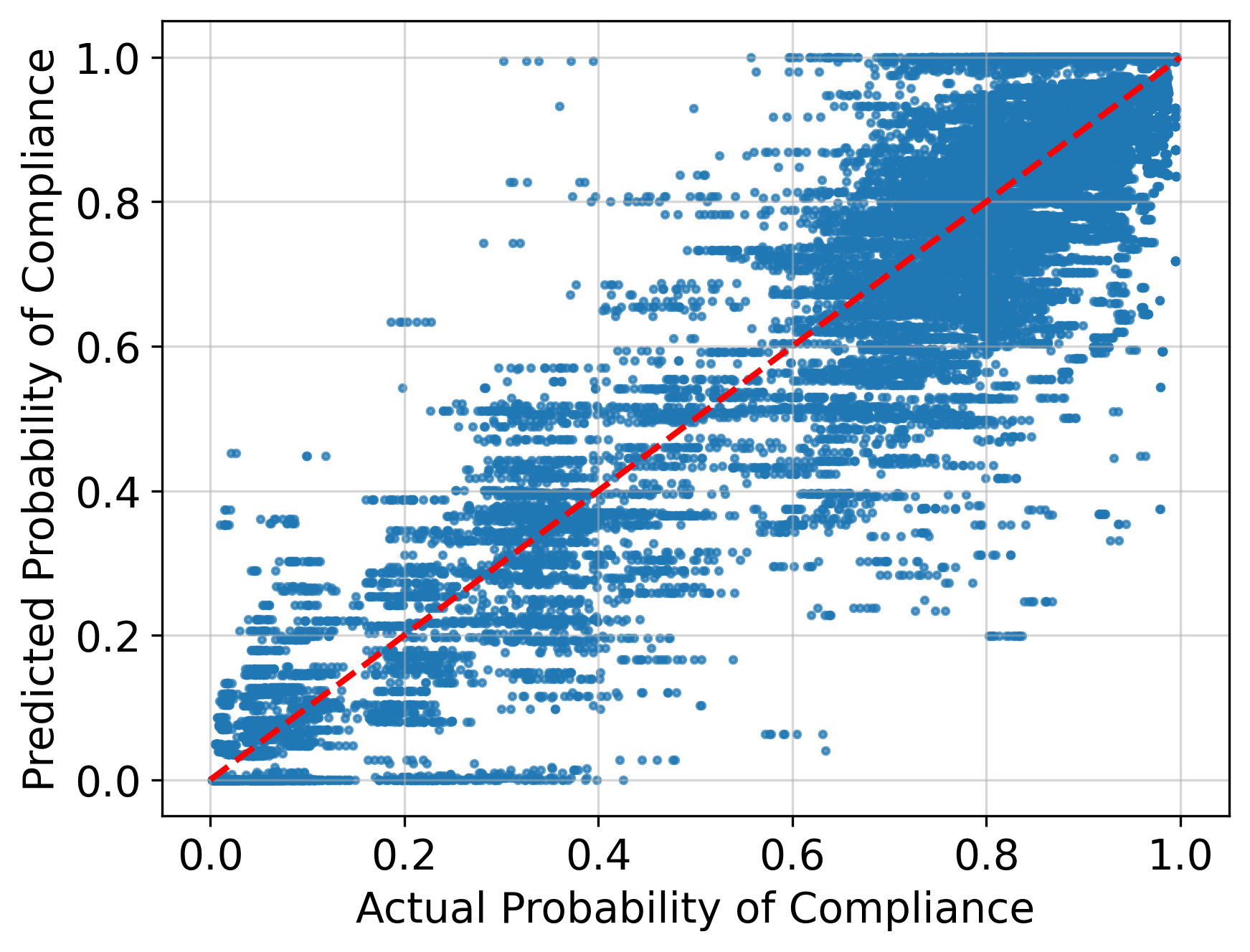}
    \caption{Comparison of the model's predicted compliance probabilities against actual observations, illustrating the strong alignment of predictions near the diagonal and indicating robust predictive accuracy.}
    \vspace{-1.2em}
    \label{fig:result-learning}
\end{figure}

\begin{figure}[!t]
    \centering
    \includegraphics[width=0.75\linewidth]{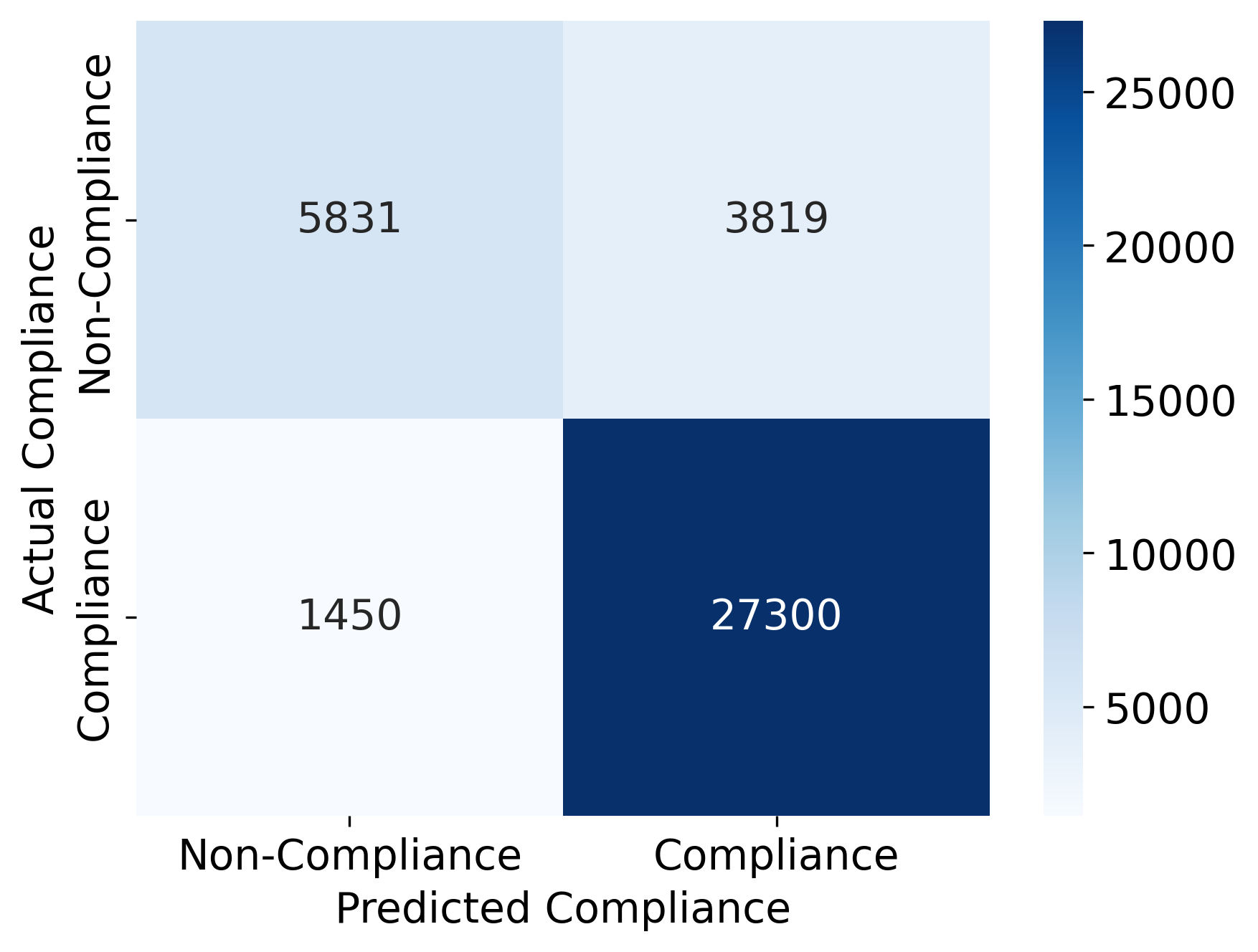}
    \caption{Confusion matrix for the compliance prediction model. The dominance of diagonal entries underscores that the model is reliably predicting compliance behavior.}
    \label{fig:result-learning-confusion}
    \vspace{-1.5em}
\end{figure}

The RF takes features such as the driver’s origin, destination, and relevant route attributes (e.g., route length) and outputs the probability of compliance with a recommended route.
Figure~\ref{fig:result-learning} illustrates the model’s performance on an unseen set by plotting actual compliance probabilities against the predicted values.
The clustered points around the diagonal line indicate that the predictions align well with the actual probability of compliance, suggesting that the learned model effectively captures the underlying decision-making patterns of travelers. 
Figure~\ref{fig:result-learning-confusion} shows the result of the compliance prediction model. Numerically, the model achieves approximately $86.28\%$ overall accuracy.
This accuracy is crucial for our subsequent route recommendation process, as it enables us to account for partial compliance in optimizing system-level traffic flows.

\subsection{Route Recommendation}

We reduce Problem~\ref{prb:assignment} to an integer programming (IP) problem that determines which route to recommend to each driver. Let $K$ be the number of possible path for each driver, i.e., $\mathcal{P}_n=\{p^1,\dots,p^K\}$, and $b^k_n\in\{0,1\}$ denote a binary variable indicating whether or not the $k$-th path for driver $n$ is recommended, i.e., $b^k_n = 1$ if $p^k$ is recommended and $b^k_n = 0$ otherwise.

\begin{problem}[Integer Programming] \label{prb:IP}
    Problem~\ref{prb:assignment} can be reformulated as the following integer program:
    \begin{equation}
        \min_{\{b_n^\mathrm{k}\}} \sum_e \Bigl|L_e^* - \sum_n\sum_p
        \sum_k b^k_n \cdot \hat{\phi}_n(p|p^k) \cdot \mathbb{I}(e\in p)\Bigr|
    \end{equation}
    \vspace{-0.5em}
    \begin{equation}
        \text{s.t. } \sum_{p^k\in\mathcal{P}_n} b^k_n = 1,~\forall n \in\mathcal{N}. \nonumber
    \end{equation}
\end{problem}
The constraint ensures that exactly one route is recommended to each driver. We employ the learned compliance model $\hat{\phi}_n$ since the true compliance behavior is generally unobservable in practical settings.
This integer programming problem can be solved using established off-the-shelf solvers such as Solving Constraint Integer Programs (SCIP) \cite{BolusaniEtal2024OO} or Gurobi \cite{gurobi}.

To evaluate the efficacy of our framework, we examine several comparative scenarios:
\textit{1) Perfect Compliance}: In the ideal case, drivers perfectly follow recommended routes. This is modeled by setting $\phi_n(p|p^k)=1$ if $p=p^k$ and $\phi_n(p|p^k)=0$ otherwise.
\textit{2) Known Compliance}: In case drivers partially comply with the recommendations, the recommender system must account for drivers' compliance patterns. We first consider the case where compliance behavior $\phi_n(p|p^k)$ is exactly known to the system operator.
\textit{3) Learned Compliance}: Our proposed approach considers compliance behavior to be unknown, and hence, utilizes learning to estimate compliance patterns $\hat{\phi}_n(p|p^k)$.
\textit{4) Naive Recommender}: Next, we consider a baseline where the system assumes perfect compliance, whereas actual drivers partially follow recommendations. In fact, this would yield the same results as applying solutions from \textit{Perfect Compliance} in reality.
\textit{5) Selfish Routing}: Finally, we consider the most basic scenario where the drivers independently select routes to minimize their own travel times without following system recommendations.
In the following subsection, we compare the performance of our approach with all the other scenarios.

\subsection{Results}

\begin{figure*}[!ht]
    \centering
    \includegraphics[width=0.83\linewidth]{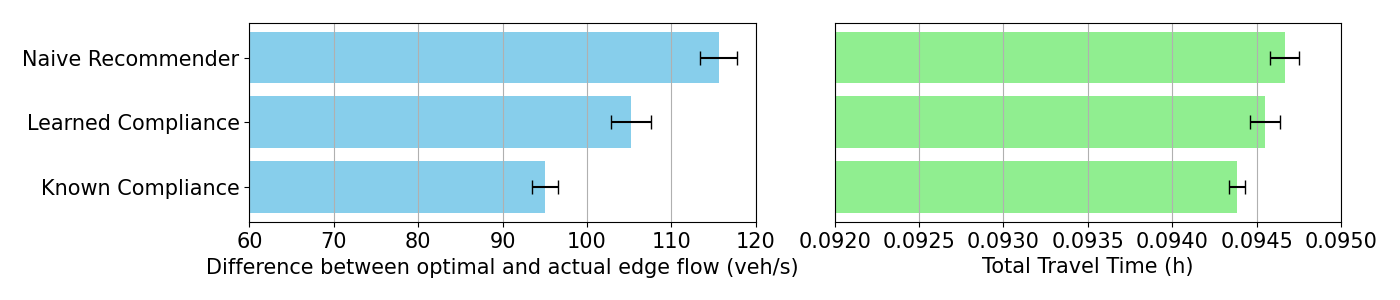}
    \caption{Comparison of different route recommendation strategies under different levels of compliance information. The left panel shows the sum of deviation from the system-optimal flow, while the right illustrates the total travel time. 
    }
    \label{fig:result-compare}
\end{figure*}

\begin{table*}[ht]
\centering
\caption{Performance comparison of various
recommendation strategies (Mean $\pm$ Standard Deviation, \textbf{bold}: our approach)}
\begin{tabular}{lcccc}
\hline
\textbf{Scenario} & \textbf{Obj. Value}  & \textbf{Flow Diff. (Optimal - Actual)} & \textbf{Total Travel Time} \\
\hline
Perfect Compliance & 13.84 $\pm$ 0.00 & 13.84 $\pm$ 0.00 & 0.0922 $\pm$ 0.0000 \\
Known Compliance & 78.23 $\pm$ 0.66 & 95.02 $\pm$ 1.59 & 0.0944 $\pm$ 0.0000 \\
\textbf{Learned Compliance} & \textbf{52.61 $\pm$ 2.96} & \textbf{105.20 $\pm$ 2.44} & \textbf{0.0945 $\pm$ 0.0001} \\
Naive Recommender & 13.84 $\pm$ 0.00& 115.63 $\pm$ 2.21 & 0.0947 $\pm$ 0.0001 \\
Selfish Routing & N/A & 912.40 $\pm$ 12.01 & 0.1631 $\pm$ 0.0046 \\
\hline
\end{tabular}
\label{tab:results}
\vspace{-1em}
\end{table*}


Given the SO flow $f_e^*$ on each edge, we derived the desired number of vehicles $L_e^*$ that should occupy each road segment using Little's Law and then solved Problem~\ref{prb:IP} to determine the optimal route recommendations considering drivers' compliance patterns.
Figure~\ref{fig:result-compare} illustrates the comparative performance of different route recommendation strategies. The results demonstrate that our compliance-aware recommendation approach consistently outperforms the naive recommender system that assumes perfect adherence. 
This performance advantage resulted from our framework's ability to anticipate and account for realistic drivers' behavior.
As expected, the known-compliance scenario, in which exact compliance probabilities are provided to the system operator, serves as a theoretical lower bound on system travel time.

Table~\ref{tab:results} provides a comprehensive numerical comparison across all five scenarios: perfect adherence, known behavior, learned behavior (our approach), naive recommender, and selfish routing.
Our method is highlighted in bold within the table to facilitate direct comparison.
The results reveal that, on average, the travel time achieved by our learned compliance approach exhibits only a $2\%$ gap from the perfect compliance scenario and $0.1\%$ gap from the known compliance case.
While the performance improvement over the naive recommender appears modest at $0.2\%$ in this particular network configuration, we anticipate that this advantage would become substantially larger in more complex networks and under more heterogeneous driver compliance patterns.
Furthermore, all recommendation strategies significantly outperform the selfish routing scenario, confirming the potential system-level benefits of intelligent route recommendations.

\section{Concluding Remarks} \label{sec:conclusion}

In this paper, we presented a new framework for route recommendations that explicitly addresses the gap between SO traffic management and driver compliance behavior. By integrating a data-driven compliance model into a stochastic optimization problem, our approach aligns recommended flows with observed decision patterns, thereby narrowing the discrepancy between theoretical and actual traffic distributions. Numerical simulations on a grid demonstrated the framework’s ability to reduce travel time compared to both naive route assignment and selfish routing baselines.

Several directions remain open for future research. First, the compliance model can adopt a more sophisticated approach, such as non-uniform deviations or context-dependent adherence, to accurately predict compliance probabilities. Second, integrating multi-modal transportation choices could extend the applicability of the framework to diverse urban mobility ecosystems. Third, exploring incentive or persuasion mechanisms alongside the learned compliance model may further encourage drivers to follow globally beneficial routes. Ultimately, this framework offers a robust and practical pathway for improving traffic congestion management in large-scale transportation systems.

\bibliographystyle{IEEEtran}
\bibliography{CPS, IDS}

\end{document}